**Comment**

# This is not about the molecules
# On the Violation of Momentum Conservation in Biology

Matthias F. Schneider

**Affiliation**: Medical and Biological Physics, Technical University Dortmund, Otto-Hahn Str. 4, 44227 Dortmund, Germany.
**Author Correspondence**: matschnei@gmail.com

**Abstract**. Conservation laws are the pillars of physics. It's what we held on to when our imagination was challenged during the days of relativity or quantum mechanics. Their violation leads to the most absurd models, so excellently exercised in the history of the perpetuum mobile. Importantly, it is *not at all sufficient* to merely accept the existence of conservation laws. Intention to obey them is required when models are developed, as conservation laws are *not* obeyed simply by accident. However evident this demand may appear, its application turns out to be quite delicate and the scientific debate of Maxwell's Demon is a beautiful demonstration of this delicateness. Here I comment on the general violation of conservation laws in most common textbook models of biological communication and outline a different route forward using the example of nerve pulse propagation.

(An extensive review on the topics mentioned here and more will appear in (Schneider, 2020))

**Universality of Conservation Laws**. Although the conservation laws of physics have certainly been accepted in biology, the conservation of momentum is so far entirely absent in most author´s minds. A glance into the most common textbooks of today's biology (Alberts et al., 1994), biochemistry (Berg et al., 2015), physiology (Guyton & Hall, 2006) or more specific literature on neurophysiology (Kandel et al., 2012) or biological signaling (Kim et al., 2015) shall suffice here as a simple demonstration. All of these books mention, in one form or another, "*the conservation of energy*" multiple times, **but not a single one has mentioned momentum conservation even once**. This is not a simple peculiarity of textbooks who are usually not designed to discuss controversies in the field. Rather, it exemplifies what is true for nearly all modern models of biological signaling or communication: ***they suffer – presumably unknowingly - from the violation of momentum conservation and the consequences of this violation can hardly be overestimated***.

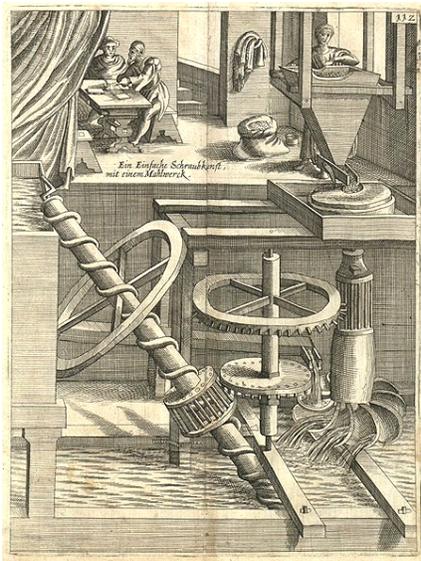

*Figure 1*. *Perpetuum Mobile Mühle. Water drops and drives a rotor to mill grains. The amount of water dropping is supposed to provide enough energy to lift the same amount of water back up and mill the seeds at the same time. It is in violation with the first law of thermodynamics.* **(Source: Deutsches Museum, Munich, Germany)**

The reason for this negligence is not discussed in detail here. It may very well origin in the tremendous success of molecular biology having an overpowering impact on the field (as opposed to physical principles). But more fundamentally it may origin in what Planck called "*the emancipation of the anthropomorphic elements*" (Planck, 1909). Simply put, this states that one of the hardest tasks in science is to step outside your own box, take the bird eye view, challenge what you have gotten used to and eliminate what is "personal" rather than objective.

As to the overpowering impact of molecular biology it is interesting to note, that in the field of nerve pulse propagation, for instance, conservation laws were well recognized (at least before the model of Hodgkin and Huxley in the 1950ties arose). Already in 1905 William Sutherland (Sutherland, 1905) for instance published a paper in *the American Journal of Physiology* proposing that nerve pulse propagation follows the same principles as sound propagation, i.e. the conservation laws of momentum and thermodynamics. An idea that has been followed by several authors (incl.

ourselves (s. below)), although still ignored (sometimes consciously) by the community following Hodgkin and Huxley's approach.

**Biology mainly consists of (hydrated) interfaces**. Physical laws are universal and apply to all systems. But what is a biological system? At least from a physical perspective one common feature throughout all biology (plants, sponges, animals, humans and all cells) is the hydrated interface. Indeed, biological systems are to a large extent networks of interfaces, which can rearrange dynamically. They are is no way simply little bags filled with bulk water (**Fig. 2**). We have therefore to apply the conservations laws (here the laws of thermodynamics and momentum conservation), to (hydrated) interfaces.

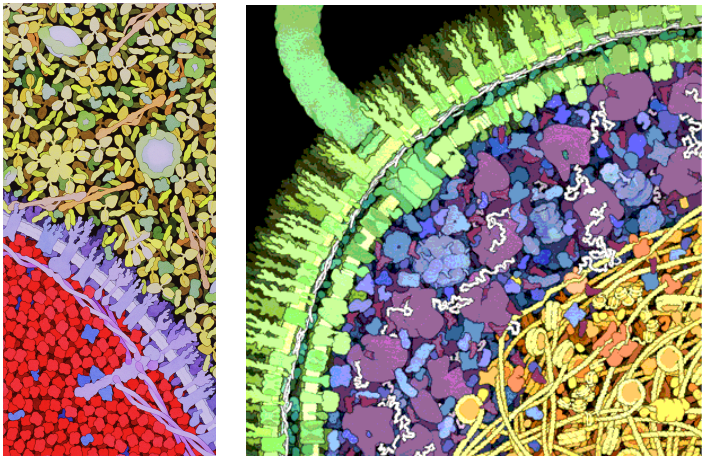

*Figure 2. An artistic illustration by David S. Goodsell of (left) a red blood cell with plasma and (right) a section of an E-Coli bacteria. The cells are crowded with molecules, which bind the intracellular water (hence interfacial or surface water). Goodsell intended to represent concentration, size and amount of cells in the correct ratio* (***Source: Scripps Research Inst., USA***).

**Interfaces have their own entropy**. Interfaces - nearly by definition - are not bulk, i.e. they are somewhat decoupled from the latter. Einstein concluded already in his first paper in 1901(Einstein, 1901), that the air-water interface of a drop of water has its own specific heat and hence its own entropy potential, i.e. an entropy potential independent of the bulk. A perturbation of such an interface is conserved. This inevitably leads – due to the conservation of momentum and the laws of thermodynamics – to propagation phenomena. The thermodynamics of the system – the state diagrams or equations of state – determines the characteristics of the propagation. Faster propagating pulses have to be expected within "hard" and slow propagation within "soft" interfaces.  Things get particularly interesting if non-linear state diagrams, as they most prominently appear near phase transitions, are present. Under such conditions, solitary waves with threshold behavior and annihilation can be excited and have been experimentally confirmed (Shrivastava et al., 2018; Shrivastava & Schneider, 2014).

Although the line of arguments applies to all hydrated interfaces (which are ubiquitous and on all scales in living systems) the most extended interface in biology is probably the axon, which can get up to several meters long depending on the species (single sensory neurons in the sauropods have been estimated to have reached a length of 40-50m). Further, these systems are studied extensively providing a vast amount of experimental data to falsify theoretical predictions. After some preliminary models of Sutherland in 1905 and Wilke in 1914 (Sutherland, 1905; Wilke, 1912), the first thorough and extensive theoretical model of nerve pulse propagation based on conservation laws was presented by Konrad Kaufmann in 1989 (Kaufmann, 1989). Based on Kaufmann's work, Heimburg and Jackson (Heimburg & Jackson, 2005) proposed the soliton model in 2005, which stresses the role of non-linear effects during NPP. In 2009 and 2012 (Griesbauer et al., 2009, 2012) we were finally able to provide experimental evidence for the propagation of linear pulses in lipid membranes, and in 2014 (Shrivastava & Schneider, 2014) we demonstrated the existence of non-linear pulses with threshold behavior and annihilation upon collision (Shrivastava et al., 2018). We also presented a theoretical model on linear (Griesbauer et al., 2012) as well as non-linear propagation (Kappler et al., 2017), which includes viscous coupling (see (Schneider, 2020)).

However, the striking similarity between NPP and sound is just the tip of the iceberg. Interfaces are ubiquitous in biology and get "hit" by perturbations (mechanically, chemically, thermally, optically…) constantly. Muscle motion, blood pressure, breathing, food uptake and digestion, release of heat, light and noise are just a few of the sources of perturbations. Pulses, presumably linear as well as non-linear, *should be* expected and it is reckless to ignore them, in particular when it comes to propagation or communication phenomena. Momentum conservation demands such pulses and no "fancy" new mechanism is required. We therefore have to expect them to be omnipresent and revisit *all biological models* where information is transported over a distance, where momentum conservation was ignored.

Finally, when rethinking the role of momentum conservation, it is important to realize it was there from the get-go. Nature and life had to evolve in their presence learning to "deal" with them symbiotically. We can only speculate at this moment, but it seems safe to say, that we may end up with an entirely new picture of what holds the living cell together to form a functional unit.

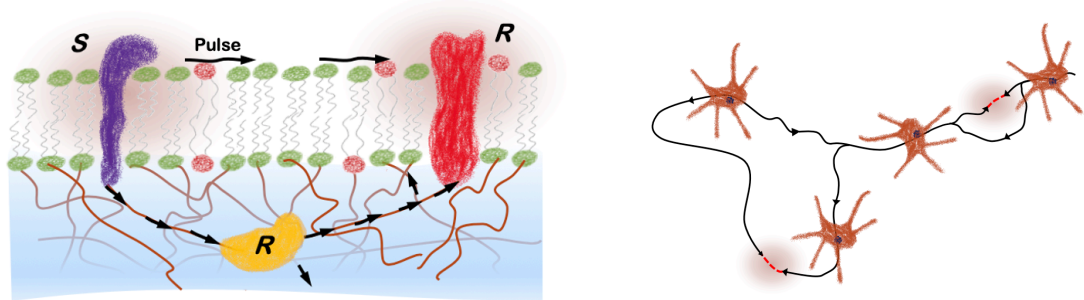

*Figure 3*. Microscopic and Macroscopic Networks of Communication. *(Left)*. A perturbation within the membrane origins from the sender S and propagates both within the cytoskeleton as well as along the membrane. The pulses eventually interact with a receiver R, which for instance could be an enzyme. This line of arguments follows from conservation laws and is of course not limited to intracellular propagation (signaling), but holds just equally within multicellular networks or organs *(right)*. A detailed review appears in (Schneider, 2020).